# SciChallenge: Using Student-Generated Content and Contests to Enhance the Interest for Science Education and Careers

**Sabri Pllana[1], Florian Huber[2], Zdenek Hrdlicka[3], Christos Mettouris[4], Asja Veber[5], Zsófia Ocsovszky[6], Chris Gary[7], Eleni Boulomyti[8], Phil Smith[9]**

[1]Linnaeus University (Sweden), [2]SYNYO (Austria), [3]University of Chemistry and Technology Prague (Czech Republic), [4]University of Cyprus (Cyprus), [5]Jožef Stefan Institute (Slovenia), [6]BioTalentum (Hungary), [7]Kinderbüro Universität Wien (Austria), [8]European Students' Union (Belgium), [9]Teacher Scientist Network (United Kingdom)

[1]sabri.pllana@lnu.se, [2]florian.huber@synyo.com, [3]zdenek.hrdlicka@vscht.cz, [4]mettour@cs.ucy.ac.cy, [5]asja.veber@ijs.si, [6]zsofia.ocsovszky@biotalentum.hu, [7]christian.gary@univie.ac.at, [8]eleni.boulomyti@esu-online.org, [9]phil.smith@nbi.ac.uk

**Abstract**

*Science education will play a vital role in shaping the present and future of modern societies. Thus, Europe needs all its talents to increase creativity and competitiveness. Young boys and girls especially have to be engaged to pursue careers in Science, Technology, Engineering and Mathematics (STEM). However, statistics still show that enrolment rates in STEM-based degree programs are decreasing. This will lead to a workforce problem in the industrial sector as well as in research and development, especially in many of the new member countries. This paper highlights a recently funded EU-research project SciChallenge (www.scichallenge.eu), which focuses on the development of novel concepts to get young people excited about science education. It uses a contest-based approach towards self-produced digital education materials from young people for young people. In cooperation with partner schools, teachers, and other youth-oriented institutions, the contest participants (individuals or groups) between the ages of 10 and 20 years generate creative digital materials (videos, slides, or infographics). The participants upload their content in social media channels and the submissions are aggregated on the SciChallenge Web Platform. The winners receive prizes funded by science-oriented industry and other stakeholders. Intelligent cross-sectoral positioning of various awareness modules on the SciChallenge Open Information Hub is expected to increase awareness on science careers. Through a strong involvement of related organizations and industries, we expect to open new opportunities for young people in regards to internships or taster days in STEM disciplines. Additionally, aggregated information on science events (such as, slams, nights, festivals) is shared. With this multi-level approach, SciChallenge may boost the attractiveness of science education and careers among young people. It is expected to improve the public engagement in science, economic prosperity and global competitiveness on a pan- European level.*

## 1. Introduction

In the 21st century, scientific and technological innovations have become increasingly important as we face the benefits and challenges of both globalization and a knowledge-based economy. According to the National Science Foundation for societies, in order to succeed in this new information-based and highly technological era, students need to develop their capabilities in Science, Technology, Engineering and Math (STEM) to levels much beyond what was considered acceptable in the past [1]. The reason is that STEM education is closely linked with economic prosperity in the modern global economy as well as with global competitiveness. Strong STEM skills are a central element of a well-rounded education and essential to effective citizenship. However, enrolment rates in STEM-based degree programs are low and leading to a workforce problem in industry. The 2009 report "STEM Supply and Demand Research" additionally states that these numbers may be 'sugar-coated': the downward trend in the numbers studying mathematics, engineering and physical sciences, is masked by a growth in the number of students for information technology (IT) and the biological sciences [2]. Vedder-Weiss and Fortus (2011) argue that the decline in motivation of adolescents to learn science is avoidable and can be addressed using appropriate pedagogical methods [3]. In order to engage student's interest in the technical career path, it is important that students establish a link between the theoretical knowledge and its application to solve real life problems early in their learning experience [4,5,6,13]. Several approaches use competitions or challenges to promote STEM subjects and discover the talent among young people, such as the International Mathematical Olympiad [7], NASA

Climate Kids & Science Fair Projects [8], Intel Global Challenge [9], BT Young Scientist & Technology Exhibition [10], SciFest [11], or EU Contest for Young Scientists [12].

In this paper, we describe the EU-funded research project SciChallenge (www.scichallenge.eu) that investigates the use of student-generated content and contests for enhancing the motivation of young people to pursue science education and careers. Using innovative digital techniques and social media, SciChallenge initializes a pan-European competition for getting young people between 10 to 20 years interested about the STEM-topics. Participants in the competition generate creative digital materials (videos, slides, or infographics) targeting one of the SciChallenge topics. SciChallenge is unique in the sense that higher-education institutions work together with knowledge-based companies, children universities, student unions and teachers of science in order to raise the interest of young people for STEM subjects. The project consortium includes nine partners from eight European countries: three higher-education institutions, three research companies and three networks (the European Children's Universities Network, the National Unions of Students in Europe, and the Teacher Scientist Network in UK).

The major contributions of SciChallenge are as follows,

- Aggregation of existing studies, good practices, projects, and toolkits on science education, scientific careers and participatory challenges and collection of educational sources in various scientific disciplines including digital sources, libraries and collaboration utilities to foster their use during the challenges.
- Elaboration of a novel concept for science challenges involving young people to produce scientific content for young people in a creative way under usage of SciChallenge Toolkits including detailed guidelines and best practices.
- Creation of a large number of SciChallenge Topic Sheets in a number of different European languages to be used as inspiration for the development of creative multimedia presentations of science topics by the young people. There is also the possibility to submit their own topics for the challenges.
- Building of a SciChallenge Web Platform with multifunctional modules including an open information hub with resource directories, a contest submission system with syndication features, dashboards as linkable hotspots with social sharing functionalities, and awareness channels.
- Spreading the idea of SciChallenge using various social media channels using the hashtag #scichallenge, distributing promotion material over the web platform and by performing innovative promotion activities targeting young people, schools and youth organisations all over Europe.
- Conducting a European-level SciChallenge with several contest categories (single, group, class) and various creative submission types (videos, infographics, slides, comics) to foster science education among young people and make them work with research knowledge resources.
- Attracting young people to scientific careers by integrating cross-sectorial awareness modules on the web platform like interactive science organisation profiles, an internship and exchange and tasters day exchange directory, and a science event navigator.

Section 2 of this paper will describe the SciChallenge approach, Section 3 will highlight implementation aspects, and Section 4 will provide some initial conclusions.

## 2. SciChallenge Approach

Figure 1 shows the general approach of the SciChallenge project. SciChallenge is tailored to the stakeholders needs and delivers a novel solution on a European level to make science education and scientific careers attractive for young people.

### 2.1 Involvement Roadmap and Fundraising Strategy

This step includes the identification of relevant challenge-related stakeholders (schools, student networks, education initiatives, parents, youth, research organisation, companies, educational NGOs) as well as the creation of an involvement roadmap and a fundraising strategy to engage the industry in supporting the challenge with prizes. An essential part is also to investigate relevant social media channels, digital tools, and science media platforms and to create multidimensional directories for the web platform.

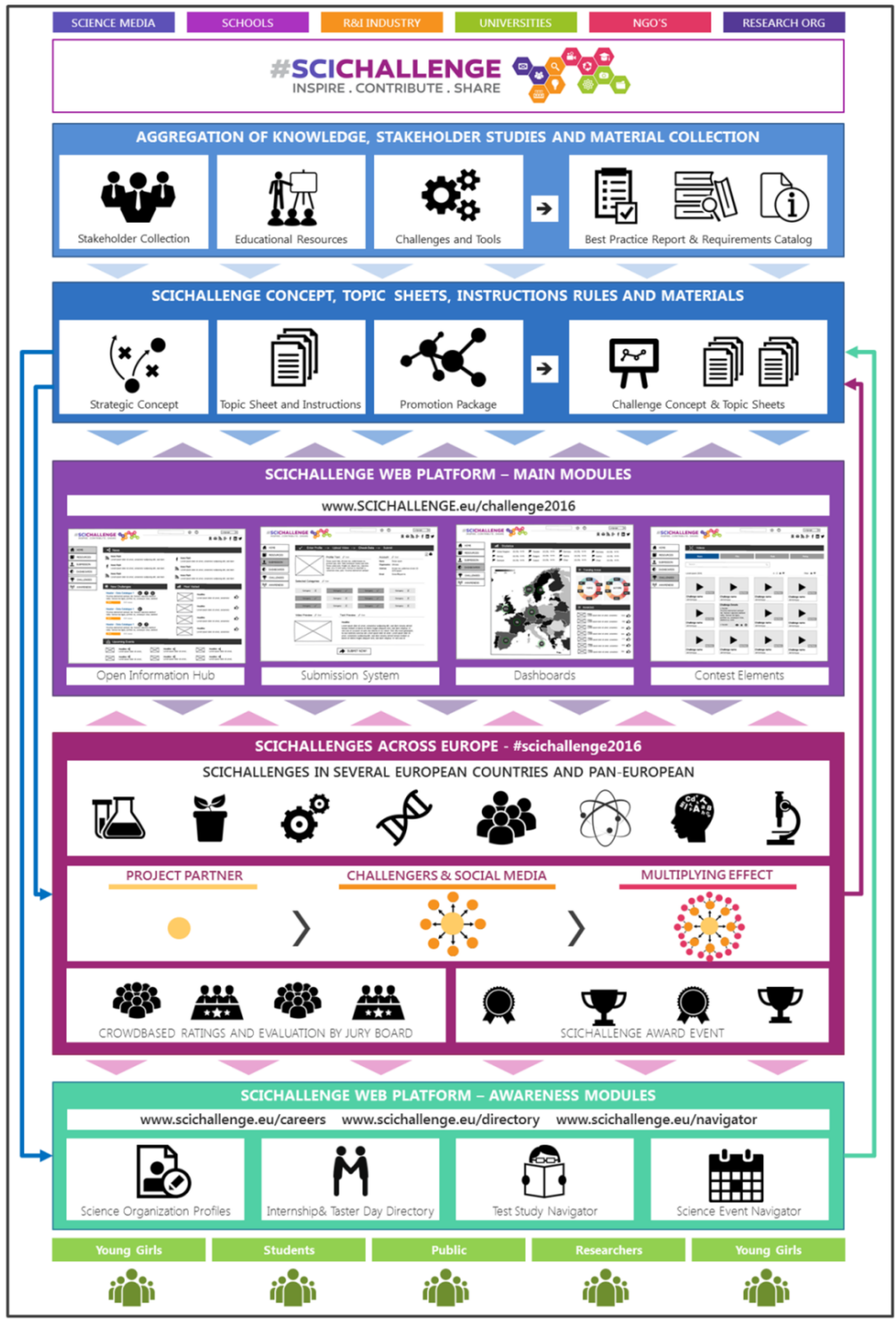

Fig. 1. SciChallenge concept.

**2.2 Web Platform**
The SciChallenge web platform (Fig. 2) consists of four parts, including an open information hub, a submission system, dashboards and awareness channels. All information and data necessary for the Dashboards and Open Information Hub are stored in a structured database. Submissions by the challenge participants are hosted on various Social Media applications to ensure effective and rapid circulation.

### 2.3 Participatory Challenge and Award Event

The core of SciChallenge is the organisation and operation of a pan-European, participatory science challenge for kids and teenagers. The pan-European nature and the opportunity to link research and commercial activities, in which multi-national working is commonplace, lift the whole concept to a new level. As a preparation step, information on research topics suitable for juveniles is prepared. This information is presented in the form of so-called topic sheets (single-page), which address teachers and pupils. They are available in English as well as in the languages of the partner countries. In addition, participants in the SciChallenge can also come up with own topics and ideas. The challenge itself closes with an Award Event taking place in Vienna, where contest winners are invited and which involves industrial sponsors, scientific organisations, and other relevant stakeholder groups. Crowd-based rating methods are developed and implemented based on Social Media channels. In addition, ratings of SciChallenge submissions by the community, an expert committee (judging board) is established.

### 2.4 Awareness Channels

The project helps to raise general awareness of STEM subjects and scientific careers by its overall project activity and the publishing of profiles of science-oriented organisations on the SciChallenge Awareness Channels. Young people also have the opportunity to access lists of internships at research or industry organisations. The general public awareness is raised by spreading information about other participatory science events such as science slams, science cafes, or science festivals as well as the communication of the project results to all affected stakeholders.

## 3. Technical Implementation Aspects

SciChallenge is designed as a highly powerful data-driven web-based platform with a flexible and scalable architecture. The architecture consists of four different layers described below

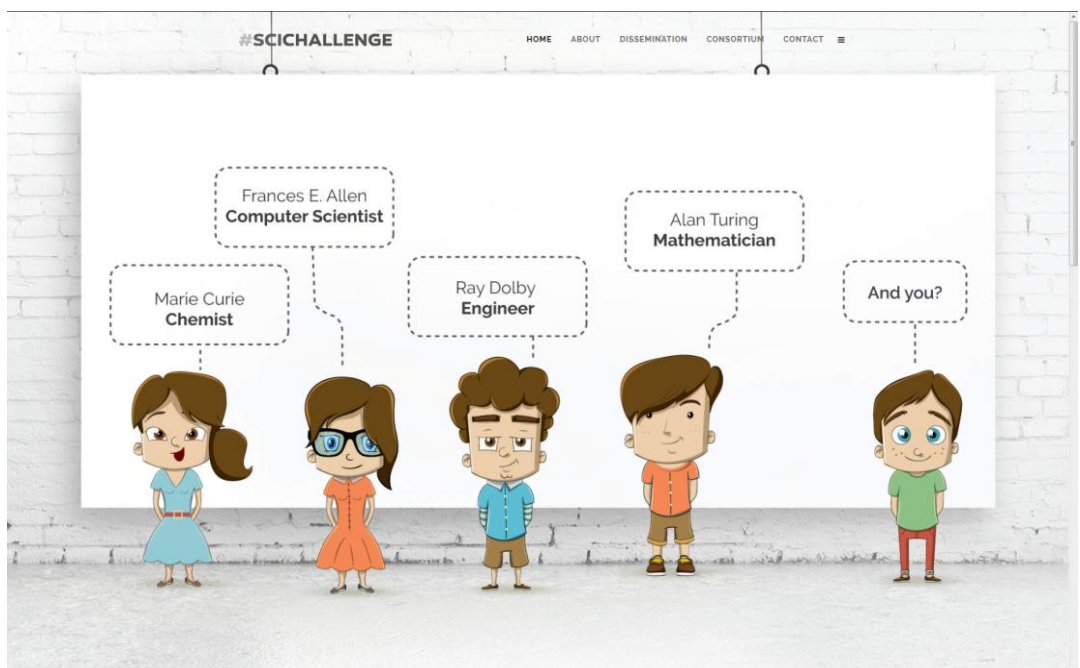

Fig. 2. SciChallenge website (www.scichallenge.eu).

### 3.1 Web Layer

The web layer (Fig. 2) comprises the following components: GUI, Process Unit, Administration Unit, and Management Unit. A navigator enables the user to obtain an overview of the challenges currently active, as obtained by the submission and presentation system. Main dashboards and visualisations will be developed using XHTML&CSS and technologies like AJAX and jQuery.

### 3.2 Contest Application Layer
This layer represents the core of the whole platform as it comprises the data for all contest participants on the one hand and the logic to process and visualise the results of the rating on the other. It comprises five components: (1) Open Information Hub, (2) Promotion Resource Library, (3) Submission and Presentation System, (4) Intelligent Rating System, and (5) Visualisation Unit.

### 3.3 Awareness Channel Layer
The awareness channel consists of four main databases, each concentrating on a different area: (1) Science Organisation Profiles, (2) Internship & Taster Day Exchange, (3) Test Study Navigator, and (4) Science Event Directory. All databases are designed in a highly scalable way by making use of modern techniques such as NoSQL databases (like mongoDB, CouchDB) or innovative search methods like Apache Solr. The content of these databases is mostly created manually and provided via the Management Unit.

### 3.4 Syndication and Sharing Layer
The syndication and sharing is responsible for promotion of the initiative as a whole by actively distributing content and information to four different groups of channels. These channels are social media, classic online media, science platforms and organization sites. Especially for social media and online media it is important to provide a consistent view, which makes it necessary to employ campaign-management tools that are also capable of distributing content to multiple channels at once.

## 4. Conclusions
SciChallenge empowers the European landscape of educational methods and resources. The project fosters the creation of creative, self-produced education materials on scientific topics by young people for young people. We hope to exploit young peoples' wish to engage digitally by combining a participatory challenge with multimedia content prepared in an appropriate platform. SciChallenge may have a long-term impact on promotion of STEM-related education and careers.

### Acknowledgements
SciChallenge has received funding from the European Union's Horizon 2020 research and innovation programme under grant agreement No 665868.